\documentstyle[aps,preprint,epsf]{revtex}
\def\myfigure#1#2{{\leftskip=0.10753\textwidth \rightskip\leftskip\small
\begin{figure}\baselineskip=14pt plus 2pt minus 1pt
\centerline{#1}\nobreak\smallskip\nobreak #2\end{figure}}}
\global\firstfigfalse

\begin{document}

\draft

\title {Kinetics of Phase Ordering in a Two-Component Fluid Membrane}
\author{P.\ B.\ Sunil Kumar and Madan Rao}
\address{Institute of Mathematical Sciences, Taramani, Madras 600 113,
India}

\date{\today}

\maketitle

\begin{abstract}
We present a novel Monte Carlo simulation of the phase separation dynamics
of a model fluid membrane. Such a phase segregation induces shape changes of
the membrane and results in local `budding' under favourable conditions.
 We present
a preliminary investigation of the nucleation regime and compute a variety of
growth laws. Our study indicates that membranes with high viscosity, 
buckle on account of phase segregation. We discuss the relation between 
this dynamical buckling and the phenomenon of `capping' seen in biological
membranes.
\end{abstract}
\bigskip
\section{INTRODUCTION}
Owing to strong hydrophobic interactions, phospholipid molecules like DMPC or
SOPC, spontaneously aggregate in an aqueous medium to form a bilayered fluid
membrane. Over length scales larger than its bilayer thickness ($\sim 50$\AA),
the membrane can be represented by a two-dimensional closed surface (vesicle)
of dimension $10-20 \mu m$, embedded in a 3-dimensional aqueous medium. In
equilibrium, the vesicle exhibits a well-defined shape and topology when subject
to constraints of fixed surface area and volume. Unlike soap films, these
equilibrium conformations are primarily determined by bending elasticity
\cite{MEMBRANE}. Shape and topology changes can be induced by changing parameters
like the temperature or osmotic conditions. Experiments like 
phase contrast video micrography 
 have established an entire taxonomy of equilibrium shapes and shape
transitions as these parameters are varied, the most dramatic being the 
`budding' transition at which a large
`parent' vesicle forms a spherical appendage attached to it by a narrow
umbilical \cite{SHAPEXP}. An extensive study of certain theoretical models
\cite{SHAPETH}
have resulted in complicated shape-phase diagrams, with shape changes occuring
at discontinuous or continuous phase boundaries. The sequence of shape changes
predicted by these theories are largely in conformity with the experiments
mentioned above.

Although the above shape changes have been observed in artificial membranes,
similar and more complex shape changes have been seen in `real' biological
membranes (e.g., plasma membrane of animal cells \cite{CELL}). 
These changes, often triggered by a change in
the chemical environment, are associated with distinct biological functions
(e.g., endocytosis, exocytosis). Unlike the simple single-component
artificial membranes discussed above, biological membranes possess around
a $1000$ different species of lipids. These lipids may differ in their
head groups, length of chains, or number and position of the unsaturated
bonds within the chain. Such multi-component lipid membranes can also exhibit
 shape and topology changes, as the area fractions of the constituent 
lipids are altered. Indeed, recent
theoretical \cite{JULICHER} and experimental \cite{DOBRY} studies on the 
shapes of
two-component artificial membranes, have shown that `budding' and even
`fission' \cite{DOBRY} may be
induced, following the phase segregation of the lipid species along the
membrane.

In this paper we argue that the time scale over which the shape of the
membrane changes, is
of the same order as the time scale over which phase segregation occurs. This
implies that the study of shape changes induced by phase separation is
a nonequilibrium problem. In the last few years there have been some
studies on the dynamical behaviour of fluid membranes (see Ref.\ \cite{CAI}
and references therein). Most of the effort has focussed on evaluating
dynamical correlation functions (and renormalized transport coefficients)
of membranes slightly away from equilibrium. On the other hand, we are
concerned with the shape dynamics of membranes far from equilibrium.
In this paper, we present preliminary results of a Monte Carlo study
of the dynamics of phase separation of a two-component lipid fluid
membrane, that incorporates the momentum density of the lipids.
As we shall see below, the relative lipid concentration is coupled
to the local curvature --- thus the shape of the membrane changes as
phase separation proceeds. Our study is restricted to the nucleation
regime, where the generic lipid concentration profile
shows patches of the minority species in the background of the other. 
Following a temperature quench, the membrane sprouts buds
locally, with the minority species forming the bud. At these early times,
coalescence does not occur. From the domain
interfacial energy density and the total mean curvature density, we extract
two length scales $R_{\phi}$ and $R_{H}$, which we show are 
related to the average 
`neck' radius and the average curvature of the bud respectively.
The length $R_{H}$, increases with time as $t^{1/4}$. Subsequent
coarsening happens with the buds coalescing or moving towards the edge
(the latter is an artifact of our `free' boundary conditions). 

We next investigate the time evolution of a single isolated patch of the
 minority species starting from a flat configuration. We find that the late time
morphology depends crucially on the in-plane viscosity of the lipids.
We find that an increase in the viscosity causes the fluid membrane
to buckle on phase segregation. This buckled shape is thus an effect
of dynamics\,; it is not an equilibrium configuration of the fluid
membrane. This buckling comes about because of the presence of
positive disclinations localised at the domain interface \cite{BUCKLING}. This
clustering induced buckling of a fluid membrane, whose in-plane
viscosity is high, might be of relevance in the phenomenon of
`capping' \cite{CELL}.

\section{Dynamics of Phase Separation of Two Component Fluids}

Let us recount that when two immiscible fluids, like water+toluene, are
cooled below their coexistence curve, they phase segregate into water-rich
 and toluene-rich regions, separated by sharp interfaces. For equal volume
 fractions
of water and toluene (critical quench), the homogenous phase is linearly unstable
to long wavelength fluctuations of the concentration variable (spinodal
decomposition). The concentration profile at intermediate times reveals
 bicontinuous regions of one fluid in the other. When the relative
concentrations of the two fluids are widely different (off-critical
quench), phase separation proceeds by the nucleation of domains of the
ordered phase, which grow and coalesce in time. In both cases, the late
time dynamics is universal - the system enters a dynamical scaling regime
\cite{BRAY}, where the equal-time concentration correlation function
behaves as $g(r,t)=g(r/t^z)$, with a growth exponent $z$ which is
independent of microscopic details. The scaling form defines a
characteristic length scale $R(t)=t^z$, interpreted as the distance between
interfaces. Scaling of the correlation function implies the existence of
only one diverging length scale. The phase separation dynamics of a binary
fluid is described by a (conserved) concentration density $\phi({\bf r},t)$
coupled to a (conserved) momentum density $\pi({\bf r},t)$. A variety of
theoretical and numerical techniques \cite{BRAY} have been employed to
obtain the form of the scaling function $g(r/t^z)$ and the exponent $z$.
Although there has been some controversy in two dimensions, our recent
Monte Carlo simulation and theoretical analyses \cite{SUNMAD} show that
there is an extended intermediate regime (for high viscosity fluids) when
$z=1/2$, with a crossover to $z=2/3$ at late times.  \section{Two Component
Fluid Membranes}

Biological membranes generically possess several ($\sim 1000$) different
lipid species, and may exhibit complex shapes and shape transformations as
a function of chemical environment or concentration of the lipid species.
Inspite of their obvious interest, a systematic experimental investigation
of the shapes of artificial multi-component membranes has not been carried
out. Let us for the moment, consider an isolated bilayered fluid membrane
made up of two distinct lipid species, e.g., DMPC and SOPC . These
phospholipids, typically differ in their degree of saturation of the length
of their hydrocarbon chains (this affects the shape of the molecule), which
in turn crucially affects the transition temperature $T_m$ below which the
lipid fluid freezes into a gel state. Typical values of this {\it main
transition temperature } are $T_m$(DMPC) $\sim 23^{\circ}C$ and $T_m$(SOPC)
$\sim 5^{\circ}C$. Since we are interested in the effects of phase
separation of the lipids on the shape of the membrane in the fluid state,
we should ensure that the phase segregation temperature lies above $T_m$.
This is done by choosing lipids with short chains or with a lot of double
bonds. 

At equilibrium \cite{JULICHER,TWOCOMP}, the energy of a 2-component
vesicle, is a unique function of the concentration of lipids
$\phi(u_1,u_2)$ and the shape of the membrane (parametrized by a 3-dim
vector of the internal coordinates on the surface, ${\bf R}(u_1,u_2)$),
subject to constraints of fixed surface area and volume. At length scales
larger than the thickness of the bilayer ($d \sim 50$\AA), the shape-energy
is determined by curvature elasticity \cite{MEMBRANE}, \begin{equation}
{\cal H}_c=\frac{\kappa_c}{2}\int (H-H_0(\phi))^2 \sqrt g \,d^2u+\kappa_g
\int K \sqrt g \,d^2u\,\,, \label{eq:helfrich} \end{equation} where the
first term contains the extrinsic curvature $H$, and the second, the
intrinsic curvature $K$. The extrinsic bending modulus $\kappa _c \approx
10^{-12}$\,ergs, is around $100 k_B T_{room}$ and so thermal fluctuations
are small, on the scale of $100 \mu m$, the typical size of vesicles. At
length scales larger than the persistance length $\xi \approx d \exp (4\pi
\kappa_c/3k_BT)$, however, thermal fluctuations are violent enough to drive
the vesicle into a self-avoiding branched polymer phase. This sets the
upper length scale cutoff in the theory. Though the intrinsic bending
modulus $\kappa_g$ has not been measured for lipid membranes, fluctuations
which change the topology of the membranes are discouraged because of the
large hydrophobicity of the lipids. In principle the moduli $\kappa_c$ and
$\kappa_g$ can depend on the local concentration $\phi$. We shall ignore
this, not being so crucial, and so the second term integrates to a constant
for a closed surface. The curvature hamiltonian, Eq.\ \ref{eq:helfrich},
replaces the bilayer by a monolayer with a spontaneous curvature (indeed
the area difference between the two leaves is, to leading order in the
thickness $d$, given by $\Delta A \approx d \,H_0$). We shall comment of
the legitimacy of this model later. The shape asymmetry between the two
lipid species gives rise to an inhomogenous spontaneous curvature which
biases the sign of the local extrinsic curvature, \begin{equation}
H_0=c_{00}+c_{01}\phi \label{eq:sponta} \end{equation}

We choose the constants $c_{00}$ and $c_{01}$ (this choice reflects the
nature of the lipid molecules) in such a way that $\phi = +1$ favours a
positive local curvature and $\phi = -1$ favours zero local curvature (Fig.
1). 

\myfigure{\epsfysize.5in\epsfbox{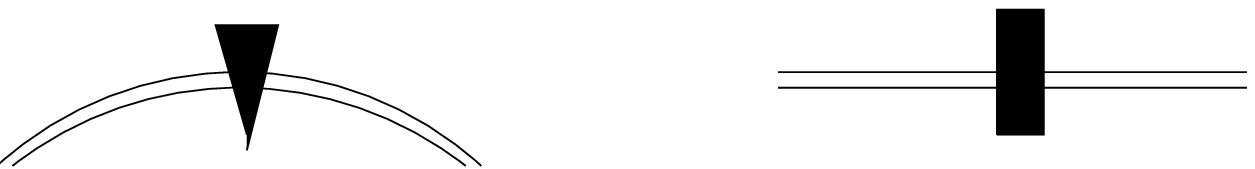}}{\vskip0.5cm FIG.~1. \ The
shape of the lipid determines the local spontaneous curvature.}

The hamiltonian describing the concentration profile of the lipids on the
membrane has the standard Landau-Ginzburg form, \begin{equation} {\cal
H}_\phi = \int \left[ \frac {\sigma}{2} (\nabla \phi)^2 - \frac {\phi^2}{2}
+ \frac {\phi^4}{4} +\mu \phi \right] \sqrt g d^2u \label{eq:landau}
\end{equation} Though the area fractions of the individual species may be
different in the two leaves of the bilayer, we shall, in the spirit of
replacing the bilayer by a monolayer with a spontaneous curvature, ignore
this fact. Given the hamiltonian, ${\cal H} = {\cal H}_c + {\cal H}_\phi $,
one can determine the equilibrium shape of the membrane subject to
constraints of constant area and volume. It was shown in Ref.\
\cite{JULICHER} that when the area fraction of one of the lipid components
is small, phase segregation results in a budding transition, with the
minority phase forming the bud. This arises becauses the interfacial
(domain wall) energy between the two species, can be made arbitrarily small
(upto the lower length scale cutoff, $d$) by forming a bud at the cost of
curvature energy. It was demonstrated, that for a wide range of parameters,
this budded configuration is favoured. Recent experiments \cite{DOBRY} have
confirmed such a shape transition in artificial membranes made of a mixture
of natural brain (sphingomyelin) lipids . 

We note that the above scenario, is valid only when phase segregation
precedes shape change, i.e., the time scale over which the local lipid
concentration relaxes is much smaller than the time scales over which shape
changes occur. In fact in membranes, both artificial and ``real'', these
relaxation time scales are comparable. The lateral diffusion coefficients
of lipids in a fluid membrane are around $10^{-8}-10^{-7}$cm$^2$/sec and so
it takes a second for a tagged lipid to traverse the entire length of the
vesicle ! Global shape changes of the membrane also occur on these time
scales. This implies that we have to address a nonequilibrium problem,
which involves the dynamics of the shape coupled to the local lipid
concentration. Indeed, it is quite plausible that the shape assumed by
membranes may be governed more by dynamics than by equilibrium physics,
much like the shapes of crystals produced from a melt. The full dynamical
equations clearly involve the following ``slow modes'', the conserved total
density of the lipids $\rho = \rho_a+\rho_b$, the conserved relative
concentration of the lipids $\phi = (\rho _a - \rho _b)/\rho$, the `broken
symmetry' shape variable ${\bf R}(u_1,u_2)$, the density of the solvent,
the momentum density of the lipid $\bf \pi$ (lipid hydrodynamics) and the
momentum density of the solvent (solvent hydrodynamics). In what follows,
we will not consider the effect of the solvent hydrodynamics --- we shall
however make precise where its inclusion would change our results. In this
paper, we will study the dynamics of the 2-component fluid membrane by a
novel Monte Carlo simulation, which incorporates the {\it overdamped
hydrodynamics} of the two-component lipid mixture.  We will not go into a
detailed justification of how our Monte Carlo method includes the momentum
density of the lipids, and refer the reader to Ref.\ \cite{SUNMAD} for a
more complete account. 

\section{Construction of a two-component fluid membrane} As a first step to
the study of the Monte Carlo dynamics of a 2-component fluid membrane, let
us learn how to ``prepare'' the system.  Our model 2-component fluid
membrane consists of two types of hard spherical beads (vertices), A and B
(the total number of A ($N_A$) and B ($N_B$) beads are fixed, $N=N_A+N_B$),
lying on a smooth, compact, 2-dim surface embedded in ${\cal R}^3$. The
beads are linked together by straight flexible tethers (bonds), in such a
way as to triangulate this compact 2-dim surface. Thus every configuration
of the surface, is represented by a {\it graph}. 
 The tethers do not intersect each other --- {\it this ensures the
planarity} of the graph. 
 The potential $V(r)$ between any two beads is infinitely repulsive at
distances less than a bead diameter $l_{min}=a$ (for all vertex pairs) and
greater than $l_{max}=\sqrt{3} a$ (for vertices connected by tethers), and
so the length of each tether can vary between these two limits. This value
of $l_{max}$ imposes self-avoidance locally \cite{MC}. We restrict the
local coordination number of every bead to lie between 3 and 9; this
ensures that the entire planar graph is {\it connected}; which forces all
the beads to lie on the 2-dim surface. A variety of ensembles suggest
themselves --- closed (spherical topology) and open (with a rigid (frame)
or free boundary). In this paper we shall study the dynamics of open
membranes with a free boundary. With this choice of ensemble, the
distribution of local coordination numbers is peaked at and symmetric about
6. We distinguish between external (edge) vertices ($V^{E}$) and internal
(bulk) vertices ($V^{I}$). Likewise an external (edge) tether ($T^{E}$)
connects two external vertices, while an internal (bulk) tether ($T^{I}$)
connects at least one internal vertex. The topology of the surface is
maintained by demanding that every internal bond $T^{I}$ is shared by two
triangles and every external bond is associated with only one triangle. 

Deformations of this fluid membrane consist of shape changes of the
membrane (particle moves), motion of particles along the membrane (fluidity
moves) and phase separation of the A and B beads (exchange moves). 

Shape changes of the membrane involve movement of the beads, which is
effected by randomly choosing a bead $i$ (of either type A or B) and then
translating $i$ to a random point (with a uniform distribution) within a
cube (of size $2l$) centered on the old position of $i$. The movement is
accepted if $(a)$ the new bond lengths lie between $l_{min}$ and $l_{max}$,
$(b)$ {\it the graph remains planar} (i.e., no bond intersections) and
$(c)$ the topology of the surface is maintained. A single Monte Carlo sweep
is $N$ attempted particle moves. These shape changes of the membrane, cost
energy ${\cal H}_c$ (Eq.\ \ref{eq:helfrich}), and so the new configurations
generated by bead movements are sampled using a Metropolis algorithm.  To
implement this, we write ${\cal H}_c$ in the following discrete form,
\begin{equation} {\cal H}_c = \kappa_c \sum_i \sum_{(ij)} \left[ H_{(ij)} +
\frac {K_i}{\sqrt 3}\right] - \kappa_0 \kappa_c \sum_{i} \,(\pm) \sqrt
{\sum_{(ij)} \left[ H_{(ij)} + \frac {K_i}{\sqrt 3}\right]}
(1+\phi_i)+{\kappa_0}^2\sum_{i} \frac {(1+\phi_i)^2}{4} a_i\,\,.
\label{eq:discrete} \end{equation} In the above equation $i$ is the vertex
index, $(ij)$ is the bond connecting the vertices $i$ and $j$ and
$\sum_{(ij)}$ is a sum over all bonds emanating from $i$. The term
$H_{(ij)} = 1-{\hat {\bf n}}_{\alpha} \cdot {\hat {\bf n}}_{\beta}$\,,
where ${\hat {\bf n}}_{\alpha}$ is the local {\it outward} normal to the
triangle labelled $\alpha$ and $(ij)$ is the common bond between the two
adjacent triangles $\alpha$ and $\beta$. In the continuum limit, however,
$H_{(ij)}$ goes over to $\frac{1}{\sqrt 3}(2H^2 - K)$ and so the term $K_i
= 2\pi - \sum \delta_{\alpha}$, which is the deficit angle at $i$, must be
added to $H_{(ij)}$ to obtain the hamiltonian, Eq.\ \ref{eq:helfrich}
(without the intrinsic curvature term), in the continuum limit.  The $\pm$
sign in the second term, denotes the sign of the local mean curvature ---
it is positive if the {\it outward} normals on adjacent triangles
(determined consistently) point away from each other and negative
otherwise. The concentration variable $\phi_i$ defined at every vertex $i$,
takes values $\pm 1$ depending on whether the vertex $i$ is occupied by an
A or a B bead. The $a_i$ in the last term, is sum of the areas of the
triangles with $i$ as the vertex. 

Motion of particles along the membrane is impeded by the tethering
constraints. To simulate fluidity, it is necessary to break-and-reconnect
bonds, in addition to moving beads.  A successful method to achieve this is
the bond reconnection algorithm \cite{MC}, which works as follows.  An
internal bond $T^{I}_{ij}$ (connecting vertices $i$ and $j$) is picked at
random. With every internal bond $T^{I}_{ij}$, we can identify two
triangles $ijv_1$ and $ijv_2$. This defines a quadrilateral with $i$ and
$j$ as one pair of opposite vertices. Clearly though $i$ and $j$ are
connected by a bond $T^{I}_{ij}$, the vertices $v_1$ and $v_2$ are not. The
bond $T^{I}_{ij}$ is now flipped, so that it connects $v_1$ and $v_2$ (the
vertices $i$ and $j$ now, do not have a tether connecting them). The bond
flip clearly changes the triangulation dynamically, and is associated with
a curvature energy cost. This flip is accepted with a Boltzmann
probability, provided the length of $T^{I}_{v_1v_2}$ satisfies the
tethering constraint, and {\it does not intersect any other tether
(planarity)}.  Let us denote the number of flip moves per Monte Carlo sweep
by $N_{flip}$. Clearly the total number of internal bonds is conserved
during this flip operation and none of the external bonds are flipped. 

The above algorithm has been employed in Ref.\ \cite{MC} to study the
equilibrium shapes of thermally fluctuating self-avoiding fluid membranes.
We have extended this algorithm \cite{SUNMAD} to study the {\it phase
ordering kinetics} of a two component fluid in two dimensions. As we have
argued in detail \cite{SUNMAD}, this Monte Carlo simulation includes the
momentum density of the fluid and so probes the late-time hydrodynamical
regime quite accurately. 
 Here we just note that a determination of the single-particle diffusion
coefficient $D$ from a computation of the ratio of the mean-square
displacement of a tagged particle to the excursion time (in MCS), shows
that $D$ increases on increasing $N_{flip}$ (Fig.\ 2). This implies that
the viscosity of the lipid can be tuned by $N_{flip}$. 

\myfigure{\epsfysize6in\epsfbox{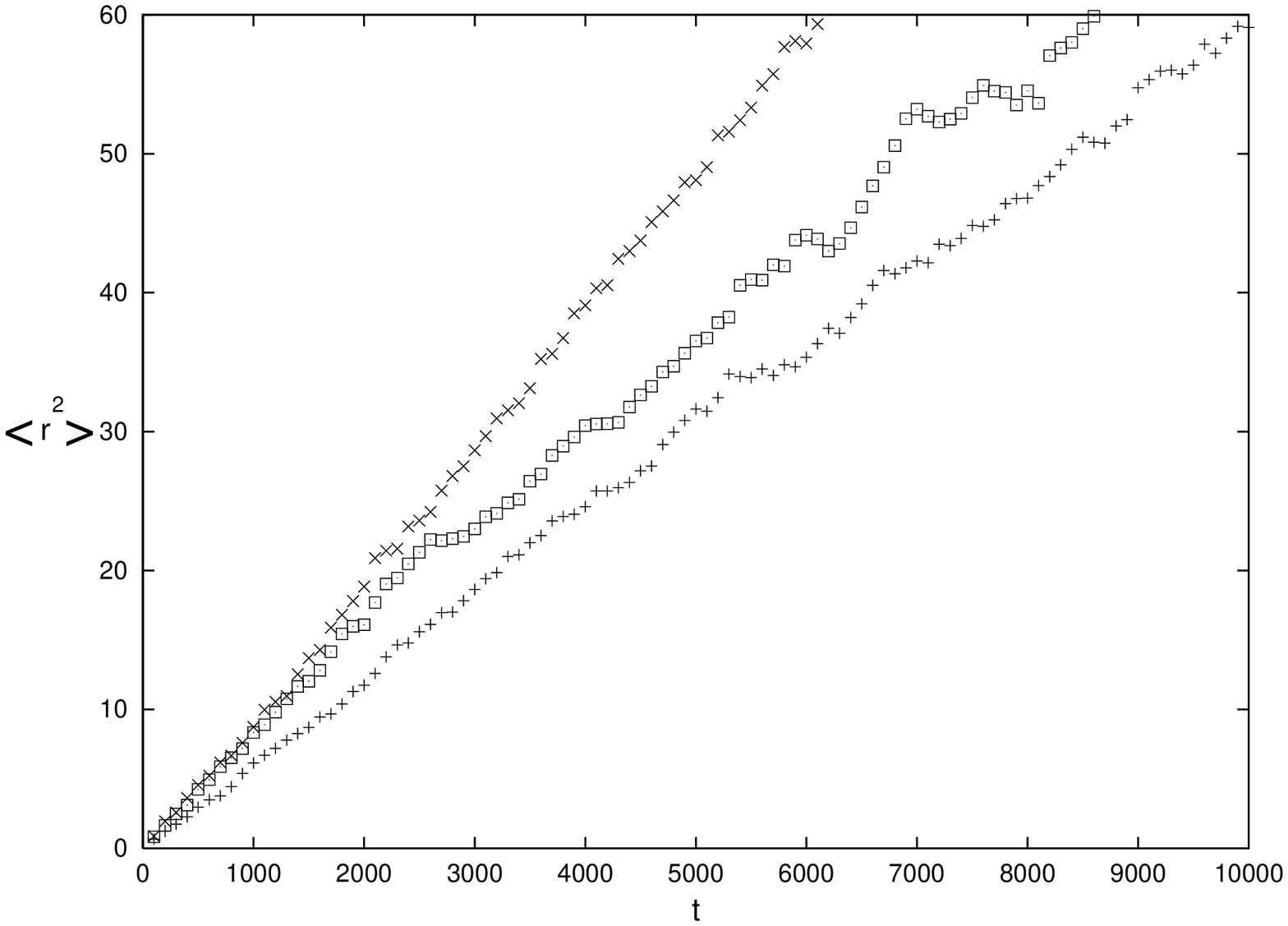}}{\vskip-4cm FIG.~2. \
Single-particle diffusion of a tagged particle for different values of
$N_{flip}$. Symbols correspond to $N_{flip}=20N$ ($\times$), $N_{flip}=10N$
($\Box$) and $N_{flip}=N$ ($+$), where $N=6000$.}

Phase separation of the two species A and B, labelled by a local
concentration $\phi_i$ ($+1$ for A and $-1$ for B), is modelled by the the
usual Kawasaki exchange dynamics \cite{BRAY} which conserves $\sum_{i}
\phi_i$. Thus locally $\phi_i$ evolves by exchanging particles at vertices
$i$ and $j$, where $j$ is connected to $i$ by a tether, with a transition
probability $W(i \leftrightarrow j)=[1-\tanh\,(\Delta E/2k_BT)]/2$, where
$\Delta E$ is the energy difference between the final and the initial
configuration, calculated from the discrete form of the hamiltonian Eq.\
\ref{eq:landau}, \begin{equation} {\cal H}_{\phi} = -\,J\,\sum_{<ij>}
\phi_{i} \phi_{j}\,\,, \label{eq:ising} \end{equation} where the sum over
$<ij>$ is over vertex pairs connected by a tether. The number of attempted
exchanges in one Monte Carlo sweep is $N_{ex}$. This completes the
discussion on the rules governing the dynamics of a 2-component fluid
membrane. 

\section{Nucleation Dynamics --- Budding and Buckling} In this paper we
shall confine our attention to the case when the concentration of A is much
smaller than B. An initial mixed phase is unstable to large amplitude,
small wavelength fluctuations of the concentration variable. This phase
separation, known as nucleation, results in isolated patches of the A-rich
phase in a background of the B-rich phase. 

We start with a flat, mixed configuration with the following parameters
$\kappa_c=3$, $k_BT=1/20$, $\kappa_0=3$ and $J=30$. This temperature is
smaller than the critical temperature of the 2-dim Ising model on a
triangular lattice, and so is unstable to the nucleation of A domains. We
display results for $N=640$, where the number of A lipids is $20$\verb+%+
of the total. The viscosity of the lipids is low\,; the number of flip
moves is $N_{flip} = 15N$ ($N_{ex}$ is maintained at $N$). The size
$R_{\phi}$ of each A patch increases as more A lipids cluster together.
When the size of the patch is such that $\kappa_c \approx 4\pi zJR_{\phi}$
(where $z$ is the average coordination number at the A$\vert$B interface),
the region locally sprouts a bud of radius $R_{H}$.  We evaluate the
interfacial energy density, $<E> = N^{-1}\sum_{<i,j>} (1-\phi_i \phi_j)$,
and the total mean curvature density $<H> = N^{-1} \sum_{i} \,(\pm) \sqrt
{\sum_{(ij)} \left[ H_{(ij)} + \frac {K_i}{\sqrt 3}\right]}$ (averaged over
several runs). It is clear that at early times, before domain coalescence
has occured, the interfacial energy density $<E> \,\approx N_{bud}
R_{\phi}/N$, while the total mean curvature density $<H> \,\approx N_{bud}
A_{bud}/NR_{H}$, where $N_{bud}$ and $A_{bud}$, the average number and area
of the buds, are constants in time. From a computation of $<H(t)>$,
averaged over 10 runs, we extract $R_{H}(t)$ (Fig.\ 3), which we fit to
$R_{H}(t) \sim A(t-t_0)^{z}$. We obtain a $z\approx0.25$ as a best fit.
This $R_{H}(t) \sim t^{1/4}$ growth can be understood from a dimensional
analysis of $\partial h/\partial t = - \nabla^4 h + \ldots$, which is a
highly simplified dynamical equation for the surface in the Monge
representation $h(x_1, x_2, t)$. This simplification ignores all the other
slow modes in the theory, and is valid at these early times (before
hydrodynamics and coalescence become prominent). 

We can parametrize the intermediate shapes of an evolving bud, from a
circular patch (disc) of radius $R_{\phi}$ to a sphere of radius $R_{H}$
(attached to the rest of the membrane by an infinitesimally narrow neck),
by spherical-cap shapes, with an area, \begin{equation} A_{bud} = 2\pi
R^2_{H} \left[\,\,1 \pm \sqrt {1-\frac{R^2_{\phi}}{R^2_{H}}}\,\,\right]\,.
\label{eq:cap} \end{equation} Thus for fixed $A_{bud}$, the two length
scales are related by $R^2_{\phi} = C_1 - C_2/R^2_{H}$, where $C_1$ and
$C_2$ are positive constants. Even if the intermediate shapes are not
exactly spherical-caps (for instance, it can be shown that these spherical
cap configurations are not Euler shapes \cite{BM}), the above relation
between $R_{H}$ and $R_{\phi}$ should still hold for arbitrary positive
constants $C_1$ and $C_2$ (Fig.\ 3, inset). 

\myfigure{\epsfysize7in\epsfbox{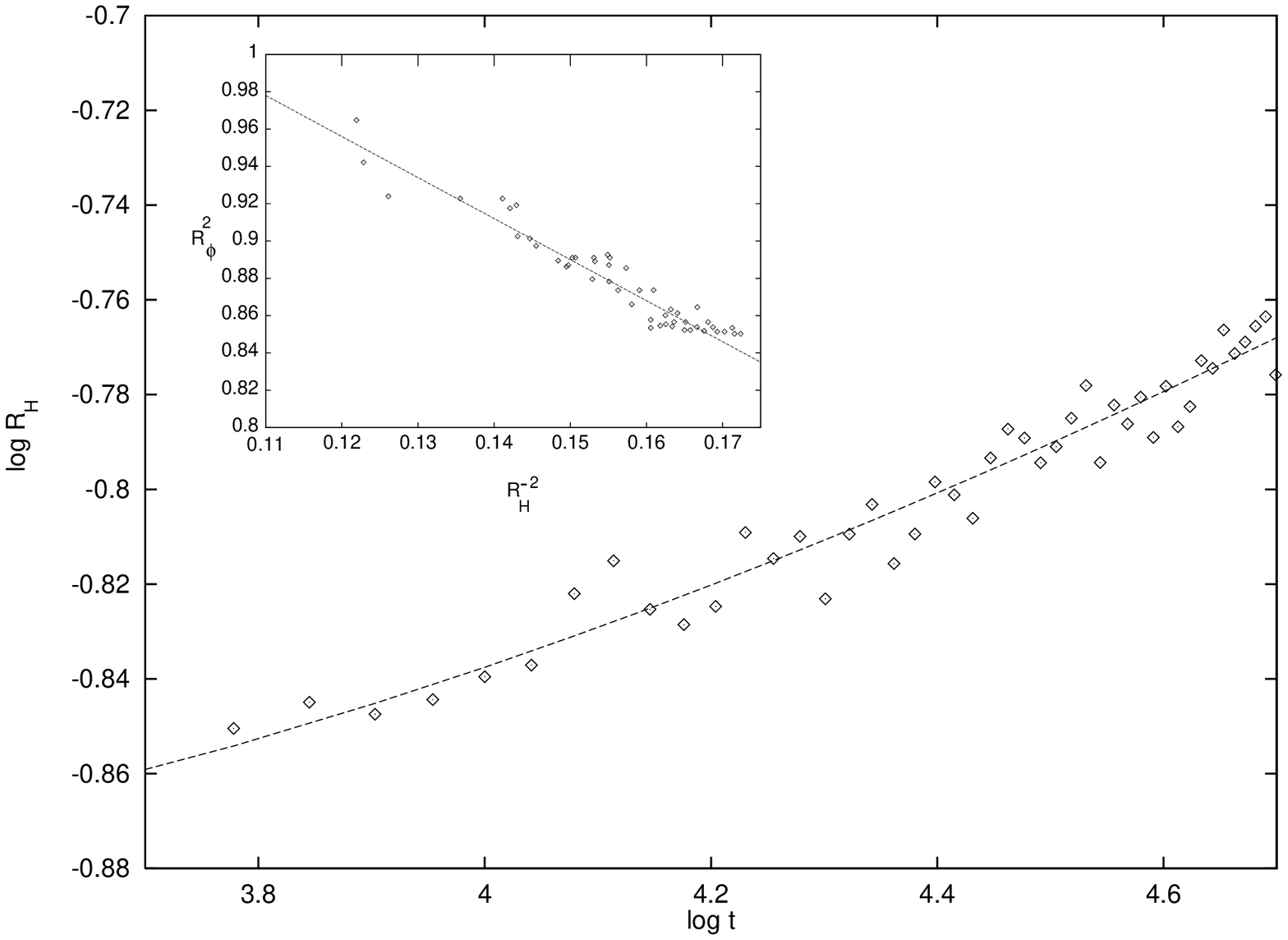}}{\vskip-6.5cm FIG.~3. \ Time
dependence of the characteristic length scale $R_{H}$, showing the
$t^{1/4}$ growth (solid line is a fit, see text).  {\it Inset} : The length
scales, $R_{\phi}$ and $R_{H}$ are related by Eq.\ 6.}

Subsequent coarsening occurs by these buds moving towards each other or
towards the boundary (artifact of our `free' boundary conditions !), and so
the late time growth laws would be drastically altered. It is at these
advanced stages in the time evolution, as buds of typical size $R_{H}$ move
towards each other, that solvent hydrodynamics would be relevant. The
motion of these buds would experience a viscous drag which would slow down
the coarsening rate. We may define two kinds of equal time concentration
correlation functions --- an Extrinsic correlation function,
 $g(r_3,t)\, \equiv \,\frac{1}{N}\,\sum_{\bf {x}}\,<\phi({\bf x}+{\bf
r_3},t)\,\phi({\bf x},t)>$ (where $\vert r_3 \vert$ is the euclidean
distance between points ${\bf x}$ and ${\bf x}+{\bf r}_3$ defined in ${\bf
R}^3$) and the Intrinsic correlation func tion $g(r_2,t) \, \equiv
\,\frac{1}{N}\,\sum_{\bf {u}}\,<\phi({\bf u}+{\bf r_2},t)\,\phi({\bf
u},t)>$ (where $\vert r_2 \vert$ is the geodesic distance between points
${\bf u}$ and ${\bf u}+{\bf r}_2$ defined on the 2-dim manifold). Light
scattering experiments would probe the fourier tranform of the extrinsic
correlation function. Which of these quantities, if any, exhibit dynamical
scaling as defined in Section II ?  It is clear that the configuration of
the membrane would deviate appreciably from flatness at late times, and so
it is likely that more than one length scale would be operative. This would
indicate a breakdown of the usual dynamical scaling at late times. We defer
an investigation of this important aspect to a later paper. 

Our next initial configuration is a flat membrane with a single patch of A
in a sea of B.  This initial condition mimics the conditions of Ref.\
\cite{JULICHER}, i.e., phase segregation is faster than shape changes.  We
know that the equilibrium shape consists of a single (or multiple) bud(s)
\cite{JULICHER}, attached to the rest of the membrane with an arbitrarily
small neck. However as we shall see now, the final shape, is crucially
dependent on the viscosity of the lipid fluid, which can be tuned by
$N_{flip}$. 

When the viscosity is low, the single patch of A sprouts out as a bud
(Fig.\ 4), and evolves steadily towards the equilibrium configuration. The
radius of the bud is related to the amount of A lipids in the patch. When
the initial configuration consists of two isolated patches of A, then two
buds sprout locally, which move towards each other (eventually they might
coalesce and form a single (or multiple) bud(s)), or towards the edge
(because of our `free' boundary condition). 

\myfigure{\epsfysize1.6in\epsfbox{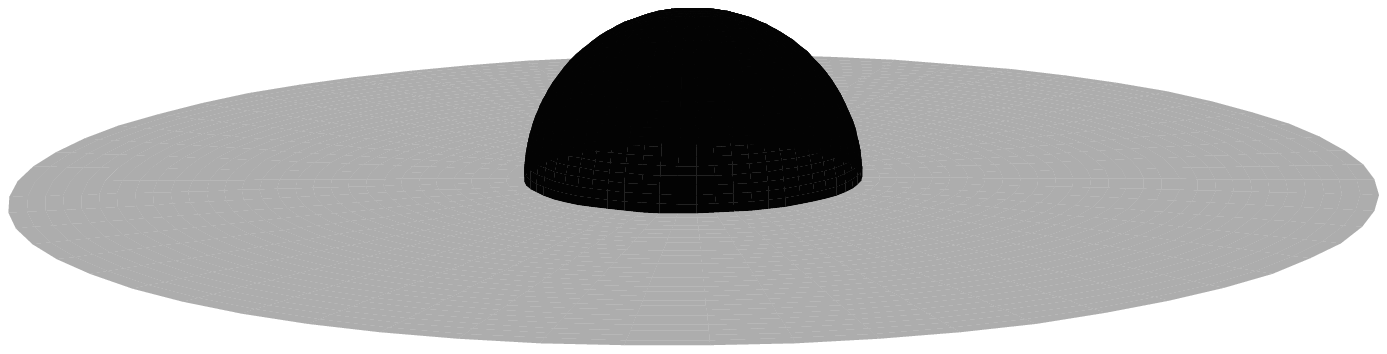}}{\vskip-0.28cm FIG.~4. \
Low viscosity, $N_{flip}=15 N$, causes Budding.}

The late time shapes are dramatically different, when the viscosities are
large. An initial configuration consisting of a single patch of A on a flat
membrane, evolves into a {\it buckled} conical shape (Fig.\ 5), with the A
species at the tip (which has positive mean curvature).  This buckled
surface is asymptotically flat. This shape is clearly not an equilibrium
configuration, but a very long-lived intermediate structure. The angle of
buckling is related to the amount of A species. 
 When the initial configuration consists of two isolated patches of A, the
late time configuration has two buckled conical structures (with the A
species at the tips), with a region of negative mean curvature in between. 

\myfigure{\epsfysize1.4in\epsfbox{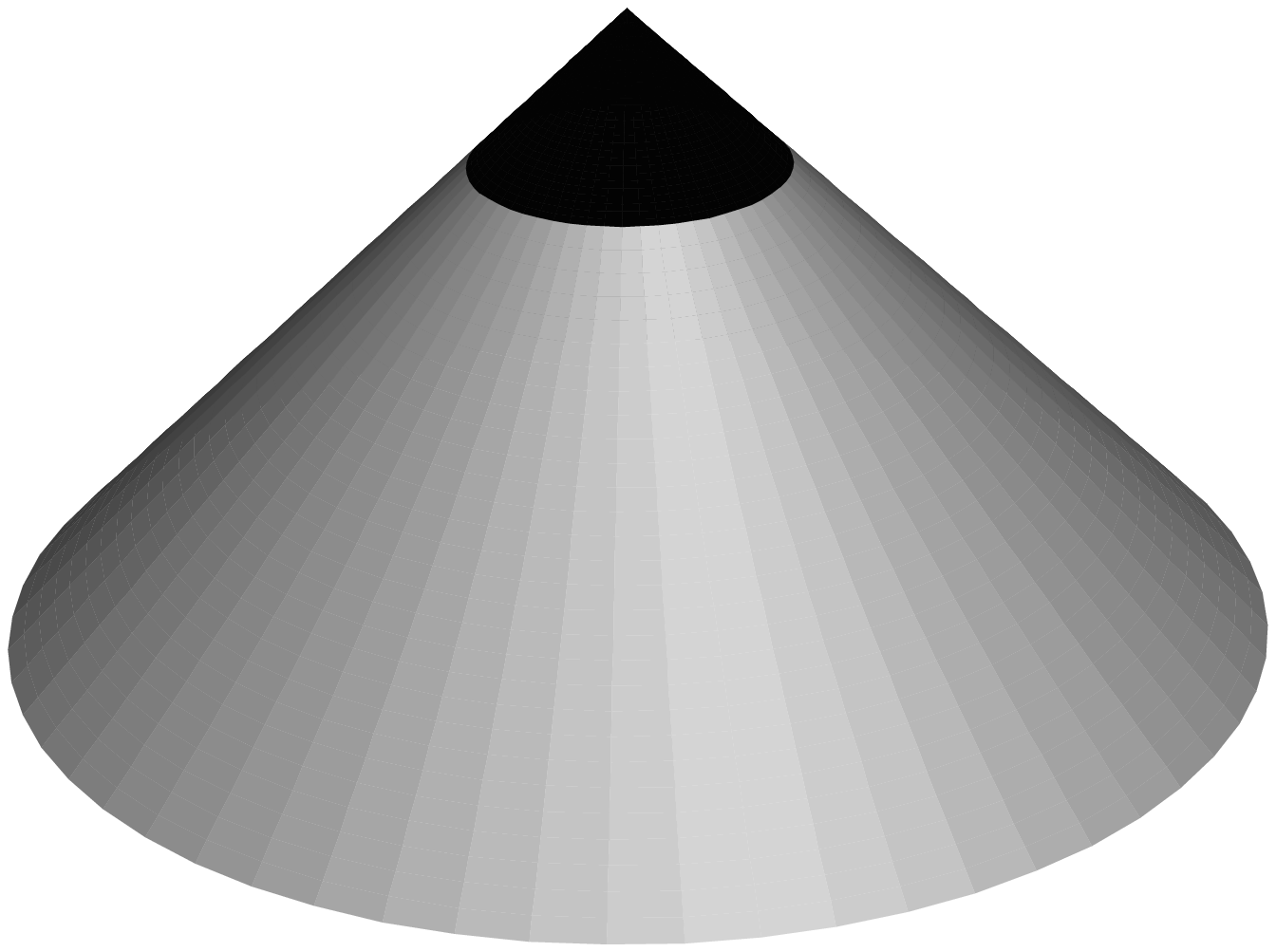}}{\vskip0.4cm FIG.~5. \
High viscosity, $N_{flip}=N$, causes Buckling.}

There is a simple reason for this dynamical buckling. The interfacial
energy can be reduced by both shrinking the interface perimeter and
lowering the local coordination number of the beads at the interface to
smaller than 6. This introduces a preponderance of positive orientational
disclinations at the interface, in an otherwise 6-fold coordinated network
\cite{NELSON}. Having forced a high viscosity on the lipids, these positive
disclinations will be relatively immobile. Further, the high viscosity
allows the shear modulus to be nonzero over small time scales (being a
fluid, it will ofcourse be zero over infinitely long time scales). Thus
over small time scales, the fluid membrane will respond elastically. As
shown in Refs. \cite{BUCKLING} an elastic membrane with a positive
disclination density $s$ (propotional to the perimeter of the A domain),
will buckle to relieve the elastic strain due to the disclinations, though
at the cost of curvature energy. The negative curvature observed in the
late-time configuration of the fluid membrane starting from an initial
configuration having two patches of A, is due to the occurence of regions
where the local coordination number is greater than 6. 

We must stress that even at these high viscosities, the membrane is fluid.
We have computed the disclination density correlation function $<n_5({\bf
u})n_5({\bf 0})>$ (where $n_5(\bf u)$ is the local density of 5-membered
rings) and find no evidence of bond orientational order. The observed
buckling is therefore truly a long-lived metastable phenomenon generated by
the slow dynamics. This dynamical buckling is reminiscent of the shapes of
crystals which are governed not by equilibrium physics but by the dynamics
of heat extraction from the melt. 

This dynamical buckling phenomenon should be observed in artifical fluid
membranes comprising of two lipids which undergo phase separation. After
the minority species have clustered to form a domain, one may induce a
photochemical crosslinking of the lipids. This would lead to an enhancement
of the in-plane viscosity $\eta$ of the membrane, by a factor, $\eta \sim
M^{3.5}$ (where $M$ is the molecular weight of the crosslinked unit
\cite{DOI}). We expect the membrane to buckle under these conditions. 

\section{Relevance to Biological Membranes} In this section we offer the
tentative suggestion that the two scenarios --- {\it budding} and {\it
buckling} --- may be observed in real biological membranes. We must stress
that the ideas expressed in this section are speculative. 

It is not clear whether, the mixture of lipids present in the plasma
membrane of cells, phase separate under normal physiological conditions.
However clustering of lipids and/or proteins can and do occur, and are
associated with a variety of biological functions. 

The relation between phase-segregation induced budding and the formation of
coated vesicles involving clathrin proteins \cite{CELL} was first pointed
out by Lipowsky \cite{JULICHER}. We note that the structure of the clathrin
coat resembles a polyhedral network of pentagons and hexagons, bound to the
vesicle. We feel that the detailed physics of clathrin coated buds, may be
more complex than the simple suggestion of phase segregation induced
budding. 

We would like to suggest that the phenomenon of `capping' observed in the
plasma membrane of cells \cite{CELL}, may be related to the dynamical
buckling discussed in the previous section. When bivalent antibodies bind
to specific plasma membrane proteins, they induce these proteins to
aggregate in patches (the membrane is still fluid). These patches collect
over one pole of the cell, which contains actin filaments, to form a `cap'.
The role of actin filaments is largely unknown \cite{CELL}, and we suggest
that it may induce partial polymerization of the lipids. This would
increase the viscosity, leading to buckling akin to the phenomenon
described above. 
 
We must caution, that any comparison with real (artificial or biological)
membranes, must take cognisance of the fact that real membranes are
bilayered having differing area fractions of the two species. Would the
qualitative features of the phase separation dynamics of such bilayered
membranes, be different from those obtained from our model ? How
independent are the lipids on the two leaves of the bilayer ? We note that
friction between the two leaves of the bilayer is extremely large $\sim
10^{7}$ dynes-sec/cm$^4$. Thus lateral diffusion of lipids on the two
leaves of the bilayer must be strongly correlated. 
 Moreover, the energy cost to change the difference in the areas $\Delta A$
of the two leaves of the bilayer is given by $(\kappa_c \pi/8 A d^2)
(\Delta A - \Delta A_0)^2$, where $A$ is the area of the vesicle. The area
difference modulus for a typical $10 \mu$m sized artificial membrane is of
the order of $10^{17}\kappa_c\,$ergs/cm$^4$ \cite{SHAPETH}. This may imply
that shape changes induced by phase segregation on the outer leaf would
pull the inner leaf along with it, to maintain the area difference between
the two leaves at a constant value $\Delta A_0$. Clearly there are a lot of
aspects to be understood in the late-time phase separation dynamics of
lipid bilayers consisting of two lipid species.

\end{document}